# Multi-photon polymerization using upconversion nanoparticles for tunable feature-size printing

[2]Qianyi Zhang, [2]Antoine Boniface, [1]Virendra K. Parashar, [1]Martin A. M. Gijs, [2]Christophe Moser

[1]Laboratory of Microsystems LMIS2, School of Engineering, Institute of Electrical and Micro Engineering, Ecole Polytechnique Fédérale de Lausanne, Lausanne, Switzerland

[2]Laboratory of Applied Photonics Device, School of Engineering, Institute of Electrical and Micro Engineering, Ecole Polytechnique Fédérale de Lausanne, Lausanne, Switzerland

**Abstract**

The recent development of light-based 3D printing technologies has marked a turning point in additive manufacturing. Through photopolymerization, liquid resins can be solidified into complex objects. Usually, the polymerization is triggered by exciting a photoinitiator with ultraviolet (UV) or blue light. In two-photon printing (TPP), the excitation is done through the non-linear absorption of two photons; it enables printing 100-nm voxels but requires expensive femtosecond lasers which strongly limits their broad dissemination. Upconversion nanoparticles (UCNPs) have recently been proposed as an alternative to TPP for photopolymerization but using continuous-wave lasers. UCNPs convert near-infrared (NIR) into visible/UV light to initiate the polymerization locally as in TPP. Here we provide a study of this multi-photon mechanism and demonstrate how the non-linearity impacts the printing process. In particular, we report on the possibility of fine-tuning the size of the printed voxel by adjusting the NIR excitation intensity. Using gelatin-based hydrogel, we are able to vary the transverse voxel size from 1.3 to 2.8 μm and the axial size from 7.7 to 59 μm by adjusting the NIR power without changing the degree of polymerization. This work opens up new opportunities for speeding up the fabrication while preserving the minimum feature size with cheap light sources.

## 1. Introduction

Three-dimensional (3D) printing, also known as additive manufacturing, allows materials to be transformed into complex objects based on computer-aided designs. Unlike traditional subtractive manufacturing such as milling and sawing, it has the advantage of rapid manufacturing of complicated structures with broad applications in research and across many industries [1]–[3]. Light-based additive manufacturing technologies driven by photopolymerization provide high resolution features (~1 μm) compared to other additive manufacturing technologies, where the feature size is limited, for example, by the nozzle size in fused deposition modeling [4] (~100 μm), or by the powder size in selective laser sintering [5] (~70-100 μm). In a photopolymerization process, a liquid mixture of monomers and photoinitiators is crosslinked upon UV or blue light excitation [6] (most common photoinitiators absorb in this spectral range). The degree of polymerization is related to the light dose, and the illuminated volume is solidified when the latter surpasses the polymerization threshold dose. Stereolithography (SLA) works by scanning a focused laser beam into a photopolymer resin to solidify the first layer of the resin [7]. Then the building platform is lowered to print the next layer and so on until it forms the desired volumetric structure. The feature size is affected by the size of the focal spot and by the diffusion of radicals. Chemical diffusion can be mitigated by adding inhibitors to the resin [8], [9]. Hence, the feature size is mainly determined by the spot size, which is on the order of the wavelength of light in the lateral (xy-plane) direction. In the axial (z) direction, absorbers can be added to reduce the layer thickness by limiting the penetration depth of light into the resin [10].

One way to improve the resolution of photopolymerization is the use of two-photon polymerization [11] (TPP). This method uses long-wavelength visible light or NIR light. The photoinitiator molecule is excited only if two photons of lower energy are absorbed at the same time. The quadratic dependence of TPP on the incident light intensity induces preferentially the reaction at the focal spot of a focused laser beam. This nonlinear absorption has a higher resolution [12] (feature size around 100 nm in the lateral direction) than single-photon polymerization and is able to create complex objects with ultraprecise features for advanced photonics [13]–[16] and nanoscale applications [17], [18]. Moreover, the cured region or voxel is highly confined in the z-direction, enabling printing in 3D without layering. The beam can also be focused at targeted points in a volume of the resin with little attenuation to achieve volumetric 3D printing [19]. However, there is always a trade-off between the printing speed and the feature size [20]. Although it is enticing to print with the minimum feature size, most objects (especially bulky structures with only a few fine features) do not need such a process. For such geometries, printing the whole piece with the smallest feature size needed is a waste of time. To speed up the printing process, one idea is to locally tune the NIR intensity (Figure 1A). Since the crosslinked voxel grows as a function of the light dose, it may result in a shorter printing time [12], [21], [22]. But, it may also induce different degrees of polymerization, hence different mechanical properties across the structure [23]. Additionally, expensive femtosecond lasers are required to achieve the high peak intensity needed for two-photon absorption. Expensive and complex instruments, together with high laser intensity [24] (~1 TW/cm$^2$), and small build volume [24] (sub-mm$^3$) prevent the broad dissemination of this technology. Recently, several printing technologies such as two-step absorption [25], [26] and triplet fusion upconversion [27], [28] have been proposed to achieve high-resolution 3D printing with continuous-wave (CW) lasers. These methods introduce nonlinear absorption into the system by utilizing real energy states in the photoinitiators or in the upconversion step before the initiation. The next generation of 3D printing systems might need nonlinear absorption in conjunction with the use of a CW light source.

Lanthanide-doped upconversion nanoparticles (UCNPs) provide another interesting solution for the nonlinear absorption with CW light [29]. A UCNP consists of a host lattice doped with activator ions (such as $Er^{3+}$, $Tm^{3+}$, and $Ho^{3+}$) that have multiple longer-lasting excitation states of similar energies and sensitizer ions ($Yb^{3+}$) with a large absorption cross-section for the NIR light. Typical UCNPs are doped with many sensitizers which absorb excitation radiation and transfer energy to neighboring activators. UCNPs sequentially absorb several incident photons of lower energy via the real metastable excited states, in contrast to two-photon absorption which involves virtual intermediate states that require simultaneous absorption. As the sequential absorption of photons does not require high intensity to increase the transition probability, photon upconversion can be achieved at an intensity of 10-100 W/cm$^2$ with an affordable CW laser [30], [31]. These UCNPs also have great photostability and a non-photoblinking nature [32]. Based on these advantages, lanthanide-doped UCNPs have already been widely used in many fields including upconversion luminescence [33], super-resolution imaging [34], photodynamic therapy [35], and other disease treatments [36]. Until now, there are only a few publications that use UCNPs in photopolymerization [30], [31], [37]–[42]. These studies reported promising results on UCNPs for 3D printing but a detailed understanding of the mechanism at stake, from the absorption of the NIR light to the polymerization of the resin, is still missing.

In this work, we propose to investigate the photopolymerization of UCNP-loaded hydrogels under NIR light exposure. The nonlinear absorption behavior of UCNPs is measured and its effect on the polymerization threshold dose and printed voxel size is studied in detail. Based on this nonlinear photosensitive material, we propose a new strategy to optimize the printing time without compromising the spatial resolution via tunable feature-size printing.

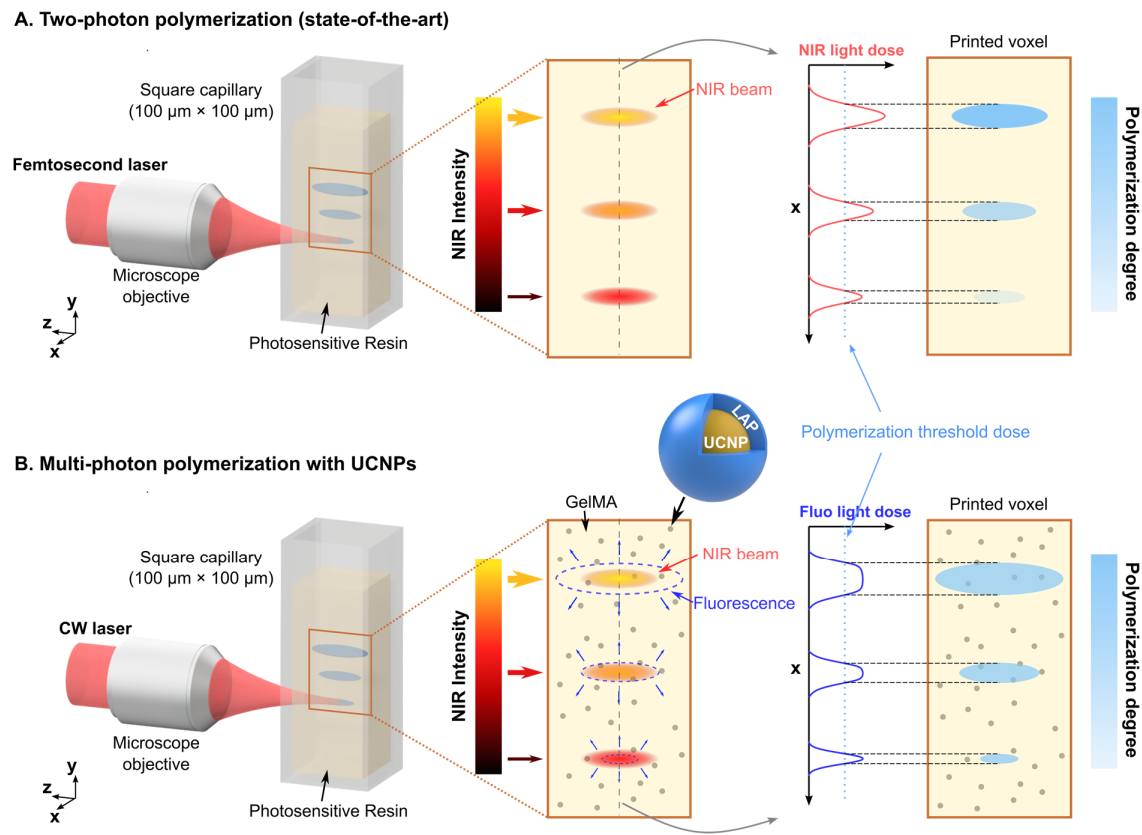

**Figure 1: Comparison between two-photon polymerization and multi-photon polymerization with UCNPs.**
(A) A femtosecond NIR beam is focused by a microscope objective through a square capillary (100 μm × 100 μm) into the resin. The printed voxel size can be tuned by adjusting the NIR intensity which produces under or over-polymerized parts, in other words, voxels with different degrees of polymerization. (B) UCNPs provide another solution for multi-photon polymerization using a CW laser. The NIR beam is focused into the photosensitive resin. The beam size of the upconverted fluorescence depends on the NIR intensity which enables printing voxels of different sizes with the same degree of polymerization.

## 2. Results

A comparison of TPP and multi-photon polymerization using UCNPs together with their strategies for tunable feature size are shown in Figure 1. 3D printing with UCNPs uses a CW laser instead of a femtosecond laser as the light source owing to the sequential absorption of NIR photons. The resin consists of UCNPs, the UV/blue light photoinitiator lithium phenyl-2,4,6-trimethylbenzoyl-phosphinate (LAP), and the commonly used hydrogel monomers gelatin methacryloyl (gelMA). The UCNPs are coated with LAP (UCNP@LAP) for better initiation efficiency. Unlike in TPP, the shape and size of the fluorescence volume vary with the NIR incident intensity. Hence, the feature size can be tuned by adjusting the NIR intensity. Additionally, one can also play with the exposure time to keep a uniform light dose to maintain the same degree of polymerization and mechanical properties across the whole print. This method allows for decoupling the relationship between the incident light dose and the degree of polymerization by providing an additional degree of freedom with the fluorescence.

The characterization of the UCNPs used in the printing process is shown in Figure 2. The core/shell NaYF$_4$:Yb$^{3+}$,Tm$^{3+}$/NaYF$_4$ nanoparticles synthesized are highly crystalline and have hexagonal morphology, with a particle size of ~60 nm (Figure 2A). The inactive NaYF$_4$ shell is ~2-3 nm to eliminate the influence of the quenching factor of the aqueous environment. A small voxel size of the fluorescence can be observed by focusing the 976-nm light beam into a medium containing UCNPs (Figure 2B), similar to TPP. To start the photopolymerization process, the wavelength of the excitation beam should fall within the absorption band of the photoinitiator. As a result, the fluorescence peak at 350 nm is the most likely one to create radicals from LAP for initiating the cross-linking (Figure 2C). Figure 2D shows the energy level diagram in NaYF$_4$: Yb$^{3+}$, Tm$^{3+}$. The Yb$^{3+}$ ion acts as the sensitizer that absorbs a NIR photon and transfers it to Tm$^{3+}$, which eventually emits photons of higher energies. The transition to a high-energy photon can be characterized by the nonlinearity parameter $n$, which is the number of NIR photons required for this transition. For example, the emission at 360 nm and 450 nm have the same $n$ value equal to 4, meaning 4 photons are needed. However, because of the saturation of the real energy state, the nonlinearity reflects the theoretical value only at low intensities, and then declines with the increased excitation intensity, resulting in a power-dependent behavior. The fluorescence spectra are measured at different 976-nm light intensities (Figure 2E) and the intensity of each peak is normalized and plotted on a log scale (Figure 2F). All the NIR intensities in this study are calculated by the power measured before the focusing objective divided by the spot area at the focal plane. We must emphasize here that the transmission of the objective (~67% at 976 nm) is not taken into account in the reported values. According to the definition, the slope of the curve in the log scale represents the nonlinearity parameter $n$. For the fluorescent peak at 350 nm, a slope of 5 is expected at low intensities and then it decreases with the NIR intensity because of the saturation of the energy states. The intensity range displayed here is limited by the sensitivity of the spectrometer. In this range, the slope decreases from 3.9 to about 0 when the light intensity varies by one order of magnitude. This power-dependent nonlinearity is crucial for having a tunable feature size of both fluorescence and printing.

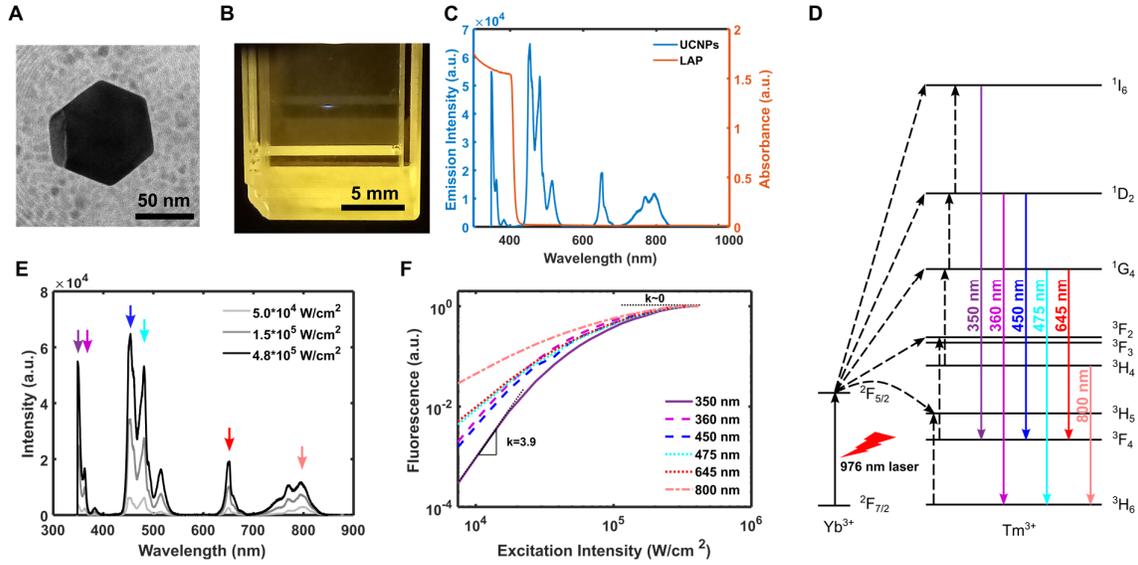

**Figure 2: Characterization of UCNPs.**
(A) Transmission electron microscopy image of a typical UCNP used in our system. (B) Fluorescent voxel in a 10 mm × 10 mm cuvette containing aqueous UCNPs produced by a CW focused laser beam (intensity at the focal plane: 19 kW/cm$^2$, wavelength: 976 nm). (C) Emission spectrum of UCNPs (blue) upon NIR light illumination and absorption spectrum of LAP (orange). a.u., arbitrary units. (D) Energy level diagram of NaYF$_4$:Yb$^{3+}$,Tm$^{3+}$. Solid arrows represent the photon absorptions or emissions, and dashed arrows represent different energy transfer processes. (E) Emission spectra of UCNPs in the gelatin solution at the excitation intensity of 50, 150 and 480 kW/cm$^2$. (F) The power-dependent emission curve at different fluorescence peaks marked in (E). Each curve is normalized to show the nonlinearity difference.

To characterize the photopolymerization dynamics of this 3D printing technology, we focus a NIR beam into the UCNP-loaded photosensitive resin contained in a 100 μm ×100 μm square capillary. We ensure the focal spot is in the center of the capillary to minimize the reflection and light-guiding effect of the glass wall. After printing, the capillary is directly imaged with an optical microscope. We avoid washing out the capillary because the gel is relatively soft and the prints are very small. Because of the low refractive index mismatch (~0.001) between the cross-linked and unpolymerized gelMA, and the small size of the voxel (1.3 μm in the xy-plane), it is very challenging to observe the printed parts with most imaging methods. Therefore, the capillary is scanned in the x-direction during printing to make the print thicker which increases the phase mismatch in the yz-plane projection and enhances the contrast for imaging it (Figure 3A). The scanning speed is 100 μm/s and the scanning range is 60 μm during all the characterization to eliminate potential effects of the scanning speed and range on the printing. The light dose is defined as the product of the NIR intensity and the illumination time. The illumination time is calculated as the product of the time per loop (a back-and-forth scanning) and the number of scanned loops. Then the light dose is only adjusted by the loop number for each intensity. Note that, since the loop number is integer or half-integer to ensure uniformity within the scanning range, the light dose can only take discrete values in our measurement setup. Then the printed object is imaged under a differential phase contrast (DPC) microscope [43] (Figure 3B). The bright and dark edges in the DPC image are specific to the distribution and the strength of the phase change (refractive index mismatch).

In this study, the bright edge represents a rise of phase change from left to right while the dark one represents a fall. The contrast (bright minus dark intensity) is related to the phase change, hence the refractive index mismatch of the object. From this DPC image, it is also possible to reconstruct a quantitative phase map of the sample, knowing some information about the illumination source [43]. The voxel sizes in y and z are characterized by their full width at half maximum (FWHM) in both directions.

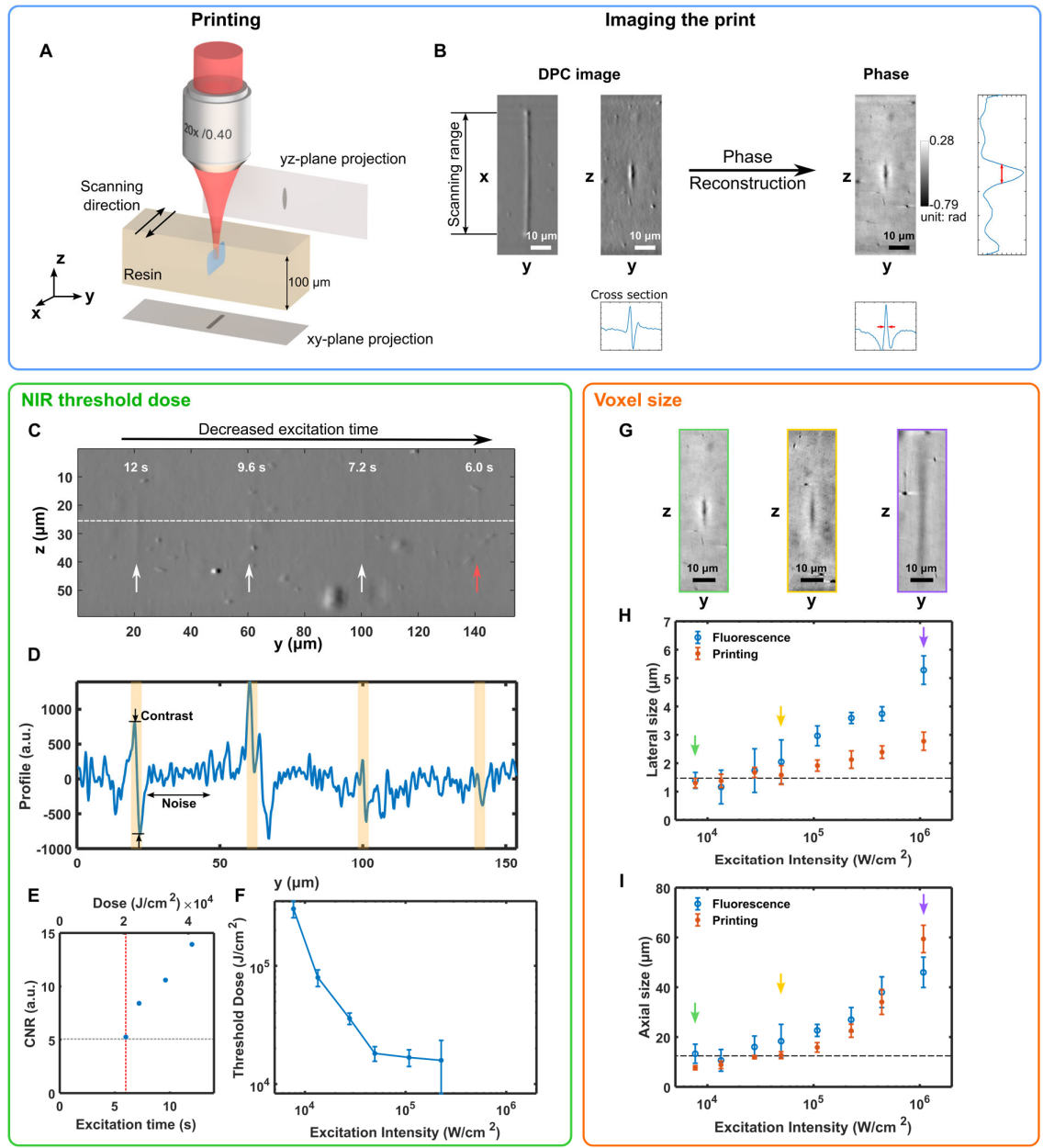

**Figure 3: Characterization of the NIR threshold dose and the printed voxel size.**
(A) Schematic illustration of the method for printing and characterizing a 'voxel'. Because of the small refractive index mismatch between crosslinked and unpolymerized gelMA, the beam is scanned on the x-axis to increase the phase change and improve the imaging contrast in the yz-plane. (B) For the characterization, the prints are imaged with a DPC microscope. The acquired images enable the reconstruction of the phase of the samples with sufficient contrast to extract features like the voxel size. (C) DPC images of voxels (marked with arrows) with different doses of NIR light (NIR intensity: $1.1 \times 10^5$ W/cm$^2$, exposure time: 6-12 s). The dose used to print the voxel indicated by the red arrow corresponds to the polymerization threshold dose at this excitation intensity. (D) Profile along y-direction across the center of the voxels (highlighted in yellow) and (E) the corresponding CNR. The gray dashed line indicates the CNR of the image we choose for the polymerization threshold, and the red dashed line indicates the threshold dose. (F) Polymerization threshold dose versus the excitation intensity. (G) Reconstructed phase of voxels printed at different excitation intensities with CNR ~15 in DPC images (for a similar refractive index change). Lateral (H) and axial (I) size of the fluorescence and the printed voxels versus the excitation intensity. The dashed line marks the focal spot size of the NIR beam in each direction.

The NIR threshold dose for the polymerization is measured from the DPC images. Figure 3C shows a DPC image of voxels printed at the NIR intensity of $1.1 \times 10^5$ W/cm$^2$ with different excitation time. Then the y-profile across the center of each voxel is extracted (Figure 3D) and the corresponding contrast-to-noise ratio (CNR) is calculated (Figure 3E). The contrast is the difference between the maximum and minimum value in the highlighted target region, while the noise is the standard deviation of the area without voxel or big dust. The polymerization threshold dose in this experiment is defined as the dose for which the CNR of the DPC image is equal to 5. In this example, the threshold marked by a red dotted line in Figure 3E correspond to an almost invisible voxel, indicated with a red arrow in Figure 3C. The light dose gradient test is performed for each NIR intensity in one capillary and the final results are obtained from 6 different capillaries (Figure 3F). The large error bar at the highest excitation intensity (i.e., corresponding to the shortest printing time) is mainly due to the discretization used for the light dose in this experiment. For the same reason, the threshold dose for higher NIR intensities cannot be measured properly because of the too slow scanning speed, which is limited by the hardware.

The voxel size is characterized from the reconstructed phase map. For voxels printed with doses near the polymerization threshold, the reconstructed phase map has very low contrast and cannot be used to accurately report the printed voxel size. Fortunately, we can assume that the voxel size does not depend on the NIR dose, at least within the range under study (see Supplementary Figure S1). Therefore, voxel images with a CNR of 15~20 are used to report on the printed voxel size for different NIR intensities. Figure 3G shows the reconstructed phase images of voxels printed at three different intensities. A significant increase in the voxel size can be seen. The NIR and fluorescent beam profiles are also measured for comparison (see Supplementary Figures S2A and B). The spot size of the fluorescence and the feature size of the resulting print both increase with the excitation intensity (Figures 3H and I), suggesting that UCNPs are responsible for the power dependency of the printed voxel size. Note that the lateral size of the printed voxel is limited by the imaging method. At low excitation intensities, the size fluctuates around 1.4 μm because of the effective pixel size (0.24 μm) and the small numerical aperture (NA 0.32). The voxel size in the z-direction decreases to the size smaller than the NIR beam with the decrease of NIR intensity, which is expected also in the y-direction. The deviation between the fluorescence profile and the resulting print in the y-direction can be explained by the method used for characterizing the fluorescence beam. The intensity beam profile is measured with a microscope objective (60X, 0.85, Newport) and a lens (f = 200 mm) with a magnification of 66.7 onto a camera. The objective scans with a z-step of 1.39 μm at the three lowest NIR intensity data points ($7.7 \times 10^3$,

$1.3×10^4$, $2.8×10^4$ W/cm$^2$) in the figure and 0.556 µm for the rest. The depth of field of this system is 1.22 µm and out-of-focus fluorescence is also collected by the camera. At low power, the fluorescent intensity drops sharply out of the focal plane because of the high nonlinearity, thus the collection out-of-focus is very deemed and its effect on the measured beam profile is negligible. At high power, the low nonlinearity (~0) creates a near-constant fluorescent intensity within the illuminated volume, thus increasing the contribution of out-of-focus fluorescence and having a stronger impact on the accuracy of the measured fluorescence profile.

Based on the nonlinearity offered by UCNP, we print a 3D structure in the form of a butterfly with tunable feature sizes (Figure 4A). The shape of the wing is simply made by adjusting the NIR intensity between $5×10^4$ and $8×10^5$ W/cm$^2$ (the capillary does not move in the z-direction). The antenna and the body require different feature sizes and are also produced by controlling the NIR intensity. DPC images of the xy-plane (Figure 4B) and the yz-plane (Figure 4C) reveal the varied feature sizes. The body and the antenna are printed with a larger scanning range, resulting in higher contrast in those regions (Figure 4C). As the wing is printed with a fixed voxel spacing in the y-direction (2 µm), voxels with a lateral size smaller than the spacing are disconnected, while voxels printed at high intensities merge and form a smooth surface. This can be improved by having an adaptive spacing based on the voxel size. Despite the large accessible range for the voxel size, a uniform phase change can be observed on the butterfly wing, indicating a uniform refractive index change, hence a uniform degree of polymerization across the structure. Note that the NIR light dose is not the same everywhere because of this power-dependent nonlinearity (as explained in Figure 3F). The total printing time of this butterfly using tunable feature size is 1 hour, while the estimated printing time at the minimum feature size would be 5 hours (estimated based on the volume). By employing this strategy, the printing time is optimized based on the structure, while preserving the spatial resolution.

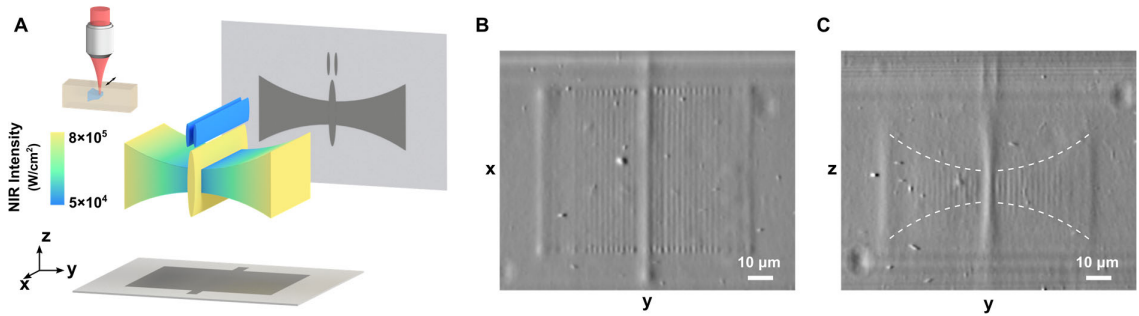

**Figure 4: Fabrication of a butterfly of tunable feature sizes with UCNP-assisted multi-photon printing.**
(A) Butterfly model and its projections in two orthogonal directions. The model is printed at different excitation intensities based on the axial feature size. The NIR light dose is adjusted for each voxel to change its size while preserving a uniform degree of polymerization across the whole structure. DPC images of xy-plane (B) and yz-plane (C) show the scanning range and the feature size, respectively. The body and the antenna of the butterfly are printed with a larger scanning range, resulting in higher contrast in the xy-plane projection.

## 3. Discussion and Conclusion

In this paper, we have characterized the polymerization dynamics of multi-photon polymerization using UCNPs. The power-dependent nonlinearity of the upconverted fluorescence is measured and its effects on the polymerization threshold dose and the obtained print feature size are investigated in detail. A clear understanding of the printing dynamics is critical to appreciate the enhancement of fidelity and resolution of additive manufacturing systems based on UCNPs, and opens the doors for its predictable and repeatable 3D printing.

Additionally, this technique also offers the possibility of changing the feature size (7.7 ~ 59 μm of the axial size, 1.3 ~ 2.8 μm of the transverse size) by modulating the excitation intensity between $7.7\times10^4$ and $1.1\times10^6$ W/cm$^2$. In contrast with TPP, here the feature size of the print can be locally adjusted without changing the degree of polymerization, ensuring uniform mechanical properties across the whole printed structure. This can be useful for applications where post-processing can be difficult or unfavorable, such as bioprinting. Various methods have been proposed to speed up the TPP process, such as multi-focus parallel fabrication [44], employing objectives with different NA [45], tuning the effective NA of the objective [46], and hybrid lithography [47]. The technique we propose does not conflict or overlap with these methods and we believe that combining them could be a way to achieve even shorter printing times.

Thanks to the high penetration depth of NIR light inside tissues, this approach should be especially relevant for bioprinting. Recent progress has already been made to prove its potential in bioprinting [31]. However, due to the low upconversion efficiency of nanoparticles, the NIR intensity needed for polymerization is much larger than the corresponding UV/blue light intensities used for single photon polymerization, posing a potential hazard to tissue because of the thermal effect. This can be improved by tuning the concentration of the doped ions [48], and the particle size [49] and optimizing the synthesis procedure [48]. In fact, a few works have already reported UCNPs for photopolymerization that uses 10~20 W/cm$^2$ of 976-nm light intensity [30], [31], [38]. The excitation intensity for photopolymerization used in this study is on the order of $10^5$ W/cm$^2$, indicating that it is not optimal.

For resins with high cell density or real tissue where light cannot be focused to a diffraction-limited spot due to the scattering, strategies based on wavefront shaping can be used to manipulate the light and refocus it through the scattering media [50]–[52], retaining the spatial resolution of the printing.

This study shows that the power-dependent nonlinearity of UCNP is at the origin of the tunable feature size for printing. This nonlinearity behavior is due to the saturation of the real energy state of this system. This technique can be combined with other one-photon polymerization systems [30] by tuning the photoinitiator concentration to print in the saturation region. The position of the saturation region can also be tuned by an order of magnitude in the intensity through adjusting the doping concentration [53]. Note that this mechanism is not unique to NaYF$_4$:Yb$^{3+}$,Tm$^{3+}$, but also appears in other UCNPs such as NaYF$_4$:Yb$^{3+}$,Er$^{3+}$ [54], and also exists in other nonlinear energy transitions accessed via real energy states such as two-step absorption [25], [26] and triplet fusion upconversion [27], [28]. By calibrating the nonlinearity and the polymerization dynamics, other upconverting processes may also be applied to tunable feature-size printing with better efficiency or higher resolution.

In summary, this study intends to provide a detailed understanding of the polymerization dynamics of the multi-photon polymerization induced by UCNPs and shows how it can be used for tunable feature-size printing. Because of the power dependency of the upconverted fluorescence, by increasing the NIR intensity from $7.7\times10^4$ to $1.1\times10^6$ W/cm$^2$, the transverse voxel size increases from 1.3 to 2.8 μm, and the axial size from 7.7 to 59 μm. This work proposes a new strategy to achieve fast additive manufacturing while preserving fine features by using power-dependent nonlinear energy transitions and paves the way for developing new techniques for printing in nonlinear photosensitive materials

## 4. Experimental section

### 4.1 Characterization

Transmission electron microscopy images were acquired on a Tecnai Osiris electron microscope. The absorption spectrum of LAP was recorded on a Lambda 365 UV/Vis spectrophotometer. Upconverted fluorescence emission spectra were measured on the setup shown in Figure S3A. The NIR and fluorescent beam profiles were characterized by the setup shown in Figure S3B. DPC images were acquired on a microscope as previously reported [43].

### 4.2 Printing

The printing was carried out using a home-built stereolithography setup shown in Figure S3C. Experimental details on the synthesis are available in Supplementary Note 4. The resin was sonicated at 40 °C for 1 min immediately after it was taken out of the fridge. Then the square capillary (100 μm ×100 μm, CM Scientific) was filled by inserting it into the resin due to the capillary action. The capillary was pasted on a glass slide on one side, which was fixed on the sample holder. The capillary was kept vertical to the ground, making it easier to view the printing process from two orthogonal directions. The position of the sample was controlled in the x, y, and z direction by three motorized stages. In the characterization of the threshold dose and voxel size, the sample was scanned in the x-direction at 100 μm/s with a scanning range of 60 μm. The spacing or voxels is 40 μm or more to eliminate the influence of the heating effect or radical diffusion from other printed voxels. A marker with a fixed excitation intensity and time was printed regularly to check the uniformity of the resin along the capillary. The printing of each excitation intensity was performed in one capillary, from higher intensities to lower ones. Then the next capillary was printed with the same procedure and served as another sample.

In the printing of the butterfly model, each wing was printed from voxels at low intensities to those of high ones, with a voxel spacing of 2 μm on the y-axis. The excitation time of each voxel was set based on its excitation intensity. The antenna and the body were printed with the lowest and highest intensity, respectively. The scanning range of the wing was 60 μm while that of the body and the antenna was 80 μm.

**Acknowledgment**

The authors thank the open-source tools (and their contributors) which were used in this work, including Inkscape.org, Python.org, and Fiji.sc.

Research funding: The authors are grateful for financial support from the Swiss National Science Foundation under project number 196971 - "Light based Volumetric printing in scattering resins."

Conflict of interest statement: The authors declare no conflicts of interest regarding this article.

# Supplementary Information

[2]Qianyi Zhang, [2]Antoine Boniface, [1]Virendra K. Parashar, [1]Martin A. M. Gijs, [2]Christophe Moser

[1]Laboratory of Microsystems LMIS2, School of Engineering, Institute of Electrical and Micro Engineering, Ecole Polytechnique Fédérale de Lausanne, Lausanne, Switzerland

[2]Laboratory of Applied Photonics Device, School of Engineering, Institute of Electrical and Micro Engineering, Ecole Polytechnique Fédérale de Lausanne, Lausanne, Switzerland

## 1. Voxel size versus NIR dose

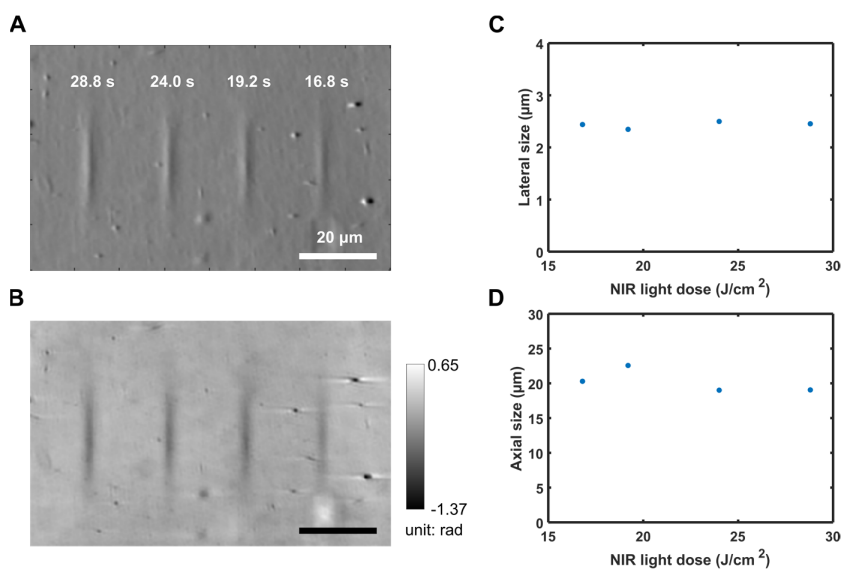

**Figure S1**
(A) DPC image of voxels with different excitation time at the NIR intensity of $1.1 \times 10^5$ W/cm$^2$ and (B) reconstructed phase of voxels printed. Lateral (C) and axial (D) size of the printed voxels versus the light dose.

## 2. NIR and fluorescence beam profile

The beam profiles of the NIR light and the upconverted fluorescence are measured by the setup in Figure S2B to understand how the tunable feature size is achieved in UCNPs-assisted 3D printing. The capillary is filled with gels containing UCNPs (10 mg/mL) and gelatin (15 wt%). The position of the capillary is adjusted so that the focal spot is in the center of the capillary, minimizing the influence of the glass wall on the recorded beam profile. The objective MO$_3$ is moved along the optical axis (z-axis) by a motorized stage and the transverse intensity profile at each *xy*-plane is recorded by a camera. The fluorescent profile of each intensity is measured at the same position of the capillary. The measurement is performed in three capillaries. Figures S2A and B show the longitudinal sections of the normalized beam profile. The NIR beam profile displays a gaussian beam with spherical aberration. Its beam size and shape do not change with the laser power. The fluorescence from UCNPs, however, has different beam profiles with the increase of the excitation intensity. For comparison, all the intensities of the fluorescence are normalized by the intensity range of the fluorescence at a NIR intensity of 8 kW/cm$^2$.

For example, the peak intensity of fluorescence at 8 kW/cm² is 1, and the peak intensity of fluorescence at 30 kW/cm² is 7, larger than the increase of the NIR intensity.

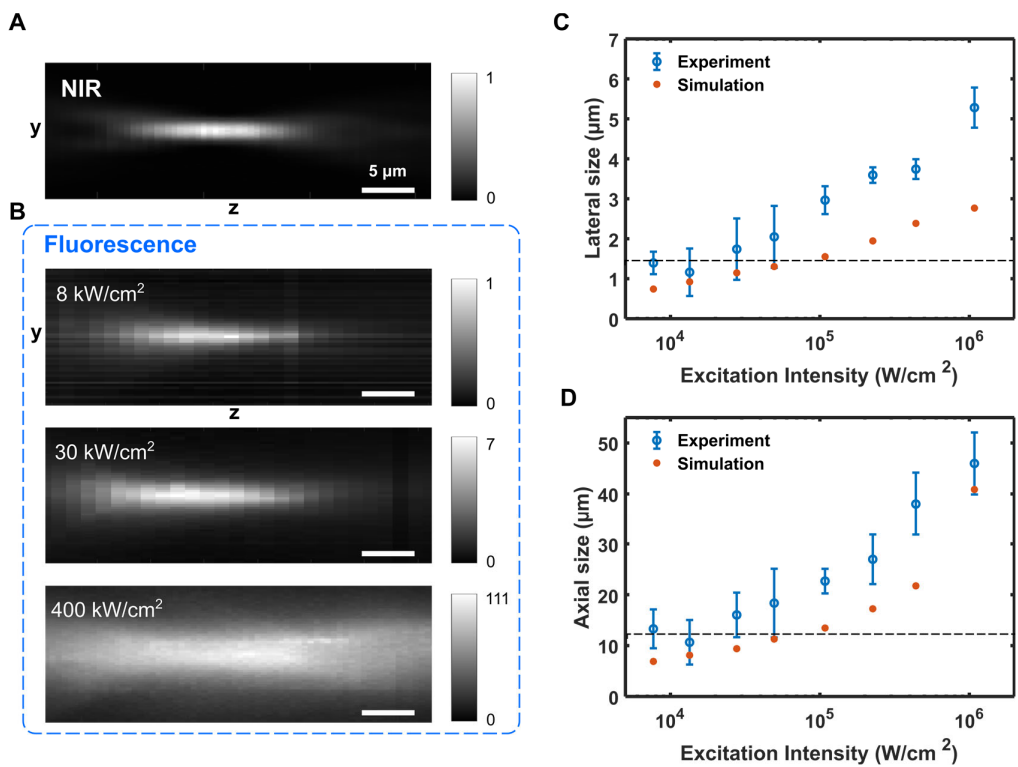

**Figure S2:**
The beam profile of the NIR light (A) and fluorescence at the NIR excitation intensity of 8, 30, 400 kW/cm². Lateral (C) and axial (D) size of the fluorescence voxel versus NIR excitation intensity. The dotted line marks the focal spot size of the NIR beam in each direction.

The lateral and axial size of the beam is defined as the FWHM of the intensity profile across the center in corresponding directions. The simulation of the fluorescent profile is based on the NIR beam profile and the power-dependent emission curve. Each pixel is assigned to a fluorescence intensity according to its NIR intensity. Then the lateral and axial size of the simulated profile is measured in the same way as that of experimental data. The discrepancies in the spot size of the experiment and the simulation are similar to those of fluorescence and printing, indicating that it is mainly caused by the way that the fluorescence profile is measured.

## 3. Setups in this study

Two different NIR beam sizes are used in this study by changing the collimator. In the measurement of emission spectra of UCNPs, a beam with constant NIR intensity (low divergence) within the gels is needed because of the power-dependent nonlinearity of the fluorescence. Therefore, a continuous-wave laser at 976 nm (900 mW, BL976-PAG900, Thorlabs) with a Polarization-Maintaining (PM) optical fiber is collimated by a lens (F230APC-980, Thorlabs, f = 4.55 mm). As the beam (d = 1 mm) does not fill the aperture of the back focal plane of the objective $MO_1$ (M Plan Apo NIR 20X, Mitutoyo), the focused beam has a beam waist of 11 mm (FWHM) with a low numerical aperture (NA ~ 0.05). The beam size is almost unchanged within the size of the capillary, providing a constant NIR intensity as well as a large volume of illumination. For printing, however, a focused beam with higher NA is preferred to have a high spatial resolution. Therefore, a large beam collimator (Thorlabs F810APC-1064) is coupled to the PM optical fiber to generate an 8-mm beam filling the whole aperture of $MO_1$. The focused beam has a diameter of 1.44 mm in the gels. The setup for measuring the beam profile also applies this optical configuration to calibrate the real beam size during the printing. The NIR intensity is tuned by modulating the current and applying different density filters (not illustrated in the figure) before $MO_1$.

Figure S3A shows the setup for emission spectra measurement. After being coupled by the collimator, the laser beam is focused by a $MO_1$ into a glass square capillary (100 μm ×100 μm, CM Scientific). The capillary is filled with gels containing UCNPs and gelatin to mimic the environment during the printing but without a chemical reaction. The fluorescence signal is then collected by a second objective $MO_2$ (UApo/340 20x, Olympus) in an orthogonal direction and coupled by a lens (LA4052-A-ML, Thorlabs) into a multi-mode fiber (QP400-025-SR/BX, Ocean Insight) and recorded by a spectrometer (Ocean Optics USB4000, Ocean Insight).

Figure S3B illustrates the setup for NIR and fluorescent beam profile measurement. The capillary is also filled with UCNPs in gelatin solution. The intensity beam profile is imaged by a microscope objective $MO_3$ (LIO-60X, Newport) and a lens (Thorlabs AC254-200-A-ML f = 200 mm) onto a camera (Basler acA1300-30gm). The objective is mounted on a motorized stage with a piezo linear actuator (Picomotor 8303, Newport) to achieve a sub-micro step size between each plane.

Figure S3C shows the setup for 3D printing. The capillary is filled with resin containing UCNP-LAP (10 mg/mL) and gelMA (15 wt%). A red light-emitting diode (LED) is applied as the illumination light source to provide a live view of the printing process in two orthogonal directions (bright-field imaging). The front view (*xy*-plane) at the center of the capillary is imaged via $MO_1$ and a lens (Thorlabs AC254-200-A-ML f = 200 mm) onto Cam 2 (Basler acA1300-30gm). NIR light is filtered by a dichroic mirror (DM, FF699-FDi01-t1-25x36, Semrock) and a band-pass (BP) filter (center wavelength = 633 mm). The side view (*yz*-plane) is collected via $MO_4$ (Olympus Plan N 20X) and a lens (f = 200 mm) and recorded by Cam 3. The NIR light is cut off by a short-pass (SP) filter (FF01-720/SP-25, Semrock).

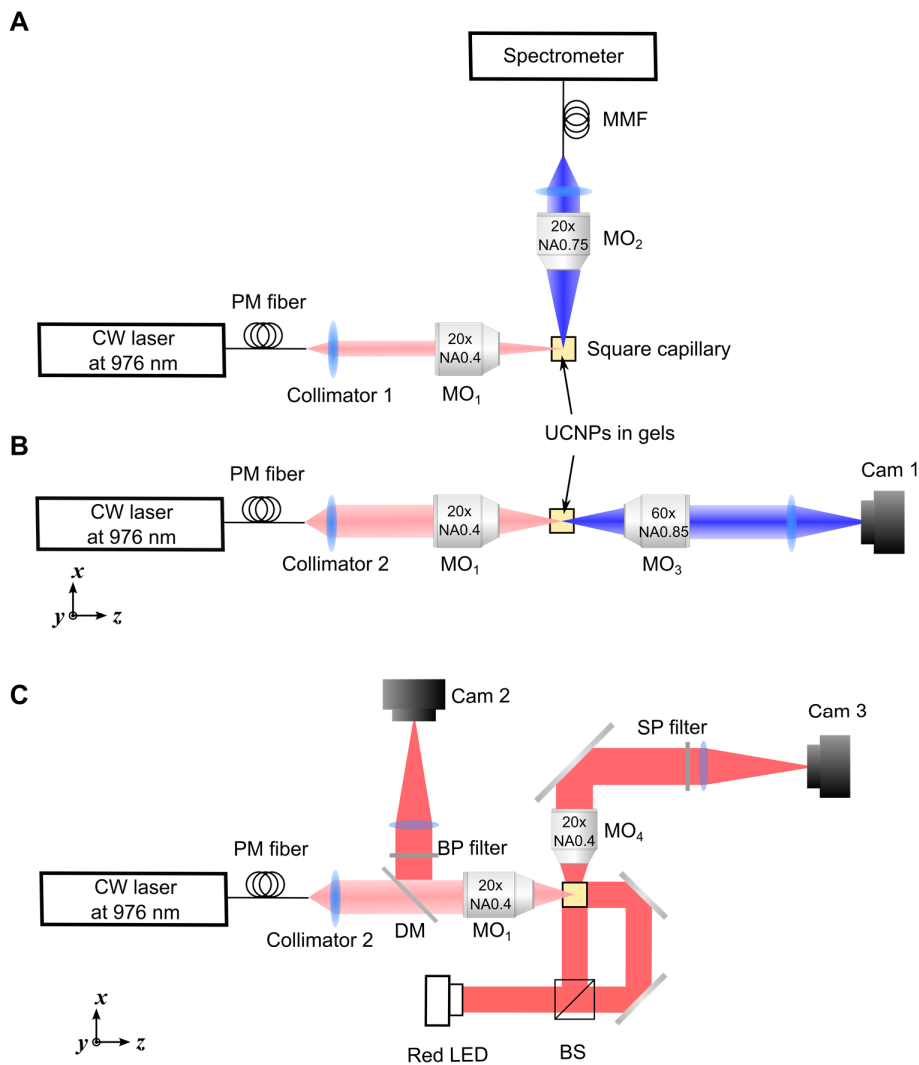

**Figure S3:**
Experimental setup. (A) Measuring the emission spectra of UCNPs. (B) Characterization of NIR and fluorescent beam profile. (C) 3D printing.

## 4. Synthesis

### Materials

Yttrium(III) acetate hydrate (99.9%), ytterbium(III) acetate hydrate (99.99+%), Thulium(III) acetate hydrate (99.9%), sodium hydroxide (reagent grade), ammonium fluoride (reagent grade), tech grade oleic acid (90%), tech grade 1-octadecene (90%), methanol, ethanol (anhydrous), cyclohexanes, hydrochloric acid (37% ultrapure), Dimethyl phenylphosphonite, 2,4,6-trimethylbenzoyl chloride, lithium bromide, 2-butanone, were purchased from Merck & Co (Sigma-Aldrich). All the solvents were deoxygenated and degassed using multiple cycles of vacuum/freeze/thawing under an argon gas atmosphere. All the syntheses were carried out under the flow of argon gas.

### Synthesis of NaYF$_4$:Yb/Tm core UCNPs (0.5 mol% Tm$^{3+}$, 30 mol% Yb$^{3+}$ doped)

In a typical synthesis, Yttrium (III) acetate hydrate (0.556 mmol), ytterbium (III) acetate hydrate (0.240 mmol), and Thulium (III) acetate hydrate (0.004 mmol) were added to a 100 ml 3 neck Schlenk flask containing oleic acid (6 ml) and 1-octadecene (15 ml) and heated to 140 C under vacuum having argon atmosphere for 90 min and cooled to 50 °C. To this, methanol solution (10 ml) of ammonium fluoride (3.2 mmol) and sodium hydroxide (2 mmol) was added dropwise and stirred for 30 min. The reaction vessel was then heated to 70 °C to remove methanol and subsequently heated to 300 °C (~10 °C/min) under argon and maintained for 60 min. The reaction mixture was then cooled to room temperature and the NCs were precipitated by the addition of ethanol, collected by centrifugation, and redispersed in cyclohexane. Repeating this process thrice, before using them as core UCNPs in the next step.

**Synthesis of NaYF$_4$ shell precursor**

Yttrium (III) acetate hydrate (0.8 mmol), was added to a 100 ml three-neck Schlenk flask containing oleic acid (6 ml) and 1-octadecene (15 ml) and heated to 140 °C under a vacuum argon atmosphere for 90 min and cooled to 50 °C. A methanol solution (10 ml) of ammonium fluoride (4 mmol) and sodium hydroxide (2.5 mmol) was added dropwise to this mixture and under continuous stirring for 30 min. The reaction vessel was then heated to 70 °C to remove methanol under argon gas and maintained for 60 min. The reaction mixture was then cooled to room temperature and used as a shell precursor.

**Synthesis of NaYF$_4$:Yb/Tm @ NaYF$_4$ core-shell UCNPs**

Layer-by-layer successive epitaxial shell growth of NaYF$_4$ was achieved on NaYF$_4$:Yb/Tm core UCNPs. Core UCNPs were added to 1-octadecene (5ml) in a 3-neck Schlenk flask and heated to 300 °C in an argon atmosphere. To this, shell precursor solution was injected @ 5 μL/sec using a nemesys syringe pump system. The ripening was done at 300 °C for 30 min. After ripening the solution was cooled down to room temperature and the core-shell NCs were precipitated and washed as outlined for core UCNPs and finally dispersed in hexane (5 ml).

**Preparation of ligand free NaYF$_4$:Yb/Tm @ NaYF$_4$ core-shell UCNPs**

The ligand-free core-shell UCNPs were prepared through the acidic treatment of the OA-capped UCNPs. In a Teflon flask, 5 mL 2M hydrochloric acid was added to 100 mg of the core-shell UCNPs, dispersed in hexane (20 mg/mL) to protonate the oleate ligands, leading to the removal of the oleic acid from the UCNPs' surface. This mixture was vigorously stirred at room temperature for 15 min and left unhindered. The organic solvent phase became transparent. The mixture was then ultrasonicated for 30 min and centrifuge at 9000 rpm for 15 min and washed with a mixture of water and ethanol (in the ratio of 1:1) twice to discard the organic layer containing the oleic acid molecules. The free UCNPs were dispersed in distilled water and stored at 4 °C.

**Synthesis of LAP photoinitiator**

LAP photoinitiator was synthesized in a two-step process. First, Dimethyl phenylphosphonite was reacted with 2,4,6-trimethylbenzoyl chloride via a Michaelis-Arbuzov reaction. At room temperature and under an argon atmosphere, 2,4,6-trimethylbenzoyl chloride (0.018 mol) was added dropwise to an equimolar amount of continuously stirred dimethyl phenylphosphonite. The reaction mixture was stirred for 18-24 hours. To this, lithium bromide (6.1 g, at least four times the amount) in 100 mL of 2-butanone was added to the reaction mixture and heated to 50 °C. After 10-15 minutes, a solid precipitate formed. The mixture was cooled to room temperature, allowed to rest for four hours, and filtered. The filtrate was washed and filtered 3 times with 2-butanone to remove unreacted lithium bromide. Excess solvent was removed by vacuum under a dark inert atmosphere. The product, lithium phenyl-2,4,6-trimethylbenzoylphosphinate (lithium acylphosphinate- LAP).

**Preparation of NaYF$_4$:Yb/Tm @ NaYF$_4$ core-shell UCNPs @ LAP photoinitiators**

UCNPs@LAP adduct was prepared by coating UCNPs (positive) and LAP (negative) through electrostatic interactions. A solution of LAP (200 mg) dissolved in Millipore deionized water was added dropwise to an aqueous solution containing 200 mg of UCNPs under ultrasonication, and the resulting mixture was further sonicated for 5-7 hours. The purified UCNP@LAP was obtained followed by centrifugation and further dispersion in water before use.

**Synthesis of gelMA**

10 g of gelatin type A (Sigma-Aldrich) was dissolved in 100 mL of phosphate-buffered saline (PBS) at 55 °C for 30 min. Then 8 mL of methacrylic anhydride (Sigma-Aldrich) was added dropwise (0.5 mL/min) and the mixture was left under stirring at 50 °C for 3 hours, followed by removal of unreacted anhydride by centrifugation at 3000 rpm for 10 min and dialysis (cellulose membrane with weight cut-off of 12–14 kDa, Sigma Aldrich) at 40 °C for 4 days against demineralized water and lyophilized. To prepare the resin for printing, 15 mg of gelMA were dissolved in 80 μL of water at 40 °C for 30 min and 20 μL of NP-LAP aqueous solution (50 mg/mL) was mixed to form a final concentration of 10 mg/mL NP-LAP and 15 wt% gelMA. A high concentration of gelMA was used to minimize the flow of the resin caused by gravity. The resin was stored at 4 °C until further use.